# Physics-Inspired Probabilistic Computing for Extremely Large-Scale MIMO Detection in Future 6G Wireless Systems

Andrea Grimaldi, *Member, IEEE*, Christian Duffee, Eleonora Raimondo, *Member, IEEE*, Edoardo Piccolo, Deborah Volpe, *Member, IEEE*, Filip B. Maciejewski, Mario Carpentieri, *Senior Member, IEEE*, Massimo Chiappini, *Member, IEEE*, Pedram Khalili Amiri, *Senior Member, IEEE*, Davide Venturelli, and Giovanni Finocchio, *Senior Member, IEEE*

***Abstract*—Extremely large-scale multiple-input multiple-output (XL-MIMO) architectures are a key enabler of forthcoming 6G wireless communication networks by allowing high data rates through massive spatial multiplexing. Here, we approach these problems with physics-inspired unconventional computing based on Ising machines (IMs). For binary modulation, probabilistic IMs (PIMs) and oscillator-based IMs achieve optimal ML detection with systems up to 2048×2048 antennas with only 100 iterations, matching optimal sphere decoder performance for computationally treatable sizes and outperforming the minimum mean-square error (MMSE) industrial standard. For M-QAM up to 256, a generalized PIM-inspired framework, based on d-dimensional probabilistic variables (p-dits) that directly encode QAM symbols, shows low bit-error-rate across sizes up to 256×256 antennas, outperforming or matching MMSE with reduced algorithmic complexity. Unlike the binary mapping, the p-dit interaction matrix is independent of the QAM order, enabling adaptive MIMO modulation. These results show a promising scalable paradigm for XL MIMO detection in future 6G networks.**

This work was supported by Project 101070287 - SWAN-on-chip - HORIZON-CL4-2021-DIGITAL-EMERGING-01, by PON Capitale Umano (CIR_00030), by Project PRIN 2020LWPKH7 “The Italian factory of micromagnetic modeling and spintronics”, by D.D. n. 47 del 20 febbraio 2025 - assunzione di ricercatori internazionali post-dottorato: CUP: D53C25000620007, and by the PETASPIN association (www.petaspin.com). The work at Northwestern University was supported by U.S. National Science Foundation 2433807 and 2322572. The work at USRA was supported by NSF CCF 1918549.
*(Corresponding author: Giovanni Finocchio.)* Andrea Grimaldi, Christian Duffee, and Eleonora Raimondo all contributed equally and are co-first authors.

Andrea Grimaldi, Edoardo Piccolo, and Mario Carpentieri are with Department of Electrical and Information Engineering, Politecnico di Bari, Bari, 70126, Italy (e-mail: andrea.grimaldi@unime.it; e.piccolo@phd.poliba.it; mario.carpentieri@poliba.it).

Christian Duffee and Pedram Khalili Amiri are with the Department of Electrical and Computer Engineering, Northwestern University, Evanston, Illinois 60208, United States of America (e-mail: christian.duffee@northwestern.edu; pedram@northwestern.edu).

Eleonora Raimondo, Deborah Volpe, and Massimo Chiappini are with the Istituto Nazionale di Geofisica e Vulcanologia, Rome, 00143, Italy (e-mail: eleonora.raimondo@unime.it; deborah.volpe@polito.it; massimo.chiappini@ingv.it).

Filip B. Maciejewski and Davide Venturelli are with the USRA Research Institute for Advanced Computer Science (RIACS), Moffett Field, California, 94035, United States of America (e-mail: fmaciejewski@usra.edu; dventurelli@usra.edu).

Giovanni Finocchio is with the Department of Mathematical and Computer Sciences, Physical Sciences and Earth Sciences, University of Messina, Messina, 98166, Italy (e-mail: gfinocchio@unime.it).

## I. INTRODUCTION

THE The performance achieved by fifth-generation (5G) wireless systems, e.g., peak data rates of 10 Gb/s in the uplink and 20 Gb/s in the downlink, end-to-end latencies below 5 ms, and reliability levels (probability of successful transmission) of 99.999%, represents a major technological milestone for modern telecommunications [1], [2]. However, future sixth-generation (6G) networks aim to substantially exceed these capabilities, targeting terabit-per-second throughput, sub-millisecond latency, and reliability approaching 99.99999% [1]. Meeting these technological requirements demands a fundamental rethinking of wireless system architectures and, critically, of the computational paradigms that underpin them. Emerging solutions include artificial intelligence (AI)-assisted channel estimation [3], advanced antenna designs [4], and new communication frameworks such as semantic communications [5]. Such approaches set the computational bottleneck associated with large-scale signal detection and remain unresolved.

A central technological pillar of 6G is expected to be the extremely large-scale multiple-input multiple-output (XL-MIMO) architecture, with antenna arrays moving beyond the 64-128 antenna regime of 5G (Massive MIMO) toward systems with 256 or more antennas. This is a requirement due to the high data rates expected for 6G operation, while also allowing narrower beamforming, more parallel data streams, and unprecedented spatial multiplexing gains. Transmissions operate using quadrature amplitude modulation (QAM), a method to encode data via signal amplitude and phase according to a symbol table (constellation), with spectrally efficient modulations such as 64-QAM and 256-QAM [2]. However, transmissions of such throughput introduce a fundamental computational challenge for detection, as described below.

Optimal detection in MIMO systems can be formulated as a maximum-likelihood (ML) estimation problem, which aims to optimize the bit-error rate (BER) of a transmission. ML MIMO is a non-deterministic polynomial-time hard (NP-hard) problem, its complexity growing exponentially with the number of transmit dimensions and the constellation size. Deterministic ML solvers such as sphere decoder [6], [7] (SD) are computationally prohibitive in the XL-MIMO regime, particularly for high-order QAM. In practice, linear ML detectors such as zero-forcing (ZF) and minimum mean-square error [8] (MMSE) are used as industrial standards due to their polynomial complexity [9]. Performance comparison with such approaches is therefore a pathway to evaluate the quality of heuristic ML estimators. Moreover, in order to be used for future 6G systems, such heuristic estimators must also satisfy stringent energy-efficiency constraints, driving a trade-off at design level between ML approximation accuracy, run time, and hardware complexity [10].

Recent advances in physics-inspired computing have opened a promising alternative in this direction. In particular, Ising machines (IMs), which are computational architectures designed to minimize a disordered, dense Ising model, provide a hardware-friendly paradigm for the solution of ML MIMO detection problems. GPU-based emulations of coherent IM dynamics have demonstrated real-time compatibility with 5G processing requirements for moderate multi-user MIMO configurations (8×8 and 16×16 antennas) [11]. Quantum reverse annealing strategies have shown improved BER performance over linear detectors in small systems (e.g., 4×6 with 16-QAM), illustrating the potential of future quantum analog hardware acceleration, assuming the resolution of operational constraints of those devices [12]. Applications of simulated bifurcation [13] and coherent IMs [14], [15] to solve small scale ML MIMO problems have also shown promising results. More recently, field programmable gate array (FPGA)-based probabilistic IMs (PIMs) with p-bits employing graph sparsification and parallel tempering have enabled 64×64 and 128×128 binary phase shift keying (BPSK) MIMO detection with an achieved full end-to-end timing evaluation of 4.8 ms [16]. These studies demonstrate that ML MIMO estimation with IMs outperforms linear detectors for small scale problems. Given that 6G networks are expected to employ antenna arrays well beyond the massive MIMO regime and to operate with high spectral efficiency through 64-QAM and 256-QAM constellations, the evaluation of IMs at the scale of these combined architectural and modulation characteristics is essential.

Here, we demonstrate that, for BPSK MIMO instances, both a p-bit PIM (bPIM) and an oscillator-based IM (OIM), with an adequate simulated annealing (SA) scheme dependent of the number of antennas, outperform MMSE for systems of up to 2048×2048 transmit/receive antennas in 100 solver iterations, approaching the theoretically optimal results achieved with SD with substantially reduced computational complexity.

However, when moving to arbitrary QAM the scalability of IMs is intrinsically limited by an error floor [13], [14] that is due to the entropic barrier present in the energy landscape arising from the exponential proliferation of metastable states [17]. This barrier is enhanced by the binary mapping of higher order modulation QAM [18]. These challenges call for innovation in physics-inspired computation to engineer a heuristic near-optimal ML detection capable of addressing next-generation 6G wireless communication architectures, thereby porting the advantages observed in moderate-scale demonstrations to XL-MIMO systems.

In this work, we found that replacing p-bits with $d$-dimensional probabilistic bits (p-dits) reduces the error floor as the physical dimensionality of the problem is preserved by the direct encoding of QAM constellation symbols without binary decomposition [19]. This suggests that the direct encoding leads to a stark decrease in the number of moves required to overcome the entropic barrier [17] or, based on the overlap gap theory [20], brings different near-optimal state space clusters closer on average. The p-dit PIM (dPIM), using replicated SA and at most 400 iterations, consistently outperforms MMSE across 4-QAM, 16-QAM, 64-QAM constellations for systems from 8×8 up to 256×256 antennas. For high-density 256-QAM constellation, the dPIM outperforms MMSE within 400 iterations for up to 64×64 antennas, and within $10^4$ iterations for 256×256 antennas. This level of performance is achieved with lower computational complexity than MMSE, with potential optimized implementations thus requiring less time to converge, making real-time implementation of XL-MIMO decoders feasible even for large antenna arrays and high modulation. The dPIM encoding is not modulation-dependent. Thus, its optimal parameters remain unchanged across the modulation order and depend only on the number of antennas. Our results establish probabilistic computing architectures with p-dits as a strategic computational paradigm for the deployment of ML XL-MIMO heuristic decoders for 6G technology.

## II. Basics of MIMO and its Mapping on the Ising Hamiltonian

Fig. 1(a) depicts a typical MIMO pipeline. The bitstream is translated into symbols according to the chosen constellation, either BPSK or a more complex $M$-QAM, with $M$ being an even power of 2. $M$-QAM is used to convert bit sequences of length $\log_2 M$ into a single symbol. Each of the possible $\log_2 M$-bit long sequences is assigned to a point of the constellation according to a minimum bit-error mapping (Gray coding), so that adjacent points of the constellation correspond to bit sequences that only differ by a single bit [21], [22].

Each instance of MIMO consists of a total of $N_\mathrm{T}$ symbols distributed across the transmitting antennas, constituting the original message $\boldsymbol{x}_0$, characterized by a power $N_\mathrm{T}E_s$, with $E_s$ being the power per symbol. This is computed as the average of the power of each constellation symbol $E_s = \sum_{x\in\mathcal{S}}|x|^2/M$, where $\mathcal{S}$ is the constellation's modulation alphabet.

The message is sent across a channel $\boldsymbol{H}$ to $N_\mathrm{R}$ receiving antennas. As the transmission is affected by noise, the received message $\boldsymbol{y}$ is given by $\boldsymbol{y} = \boldsymbol{H}\boldsymbol{x}_0 + \boldsymbol{n}$ where $\boldsymbol{n}$ is an additive white Gaussian noise (AWGN) with power $n_i$, randomly sampled from a Gaussian distribution with $\sigma^2$ variance [21]. This is defined in terms of the bit-energy-to-noise density ratio per receiver antenna $E_b/N_0$, with $\sigma^2 = N_\mathrm{T}E_s/(\log_2 M \cdot E_b/N_0)$. $N_\mathrm{R}$ should be at least equal to $N_\mathrm{T}$, and cases in which $N_\mathrm{R} > N_\mathrm{T}$ are progressively easier as $\boldsymbol{H}^\dagger\boldsymbol{H}$ is better conditioned and statistically approximates a scaled identity matrix as $N_\mathrm{R} \to \infty$ (channel hardening) [23]. In all instances considered in this work, the number of antennas $N_R \times N_T$ was chosen so that $N_R = N_T = N$, while all BER trends are expressed in terms of $E_b/N_0$ in decibel.

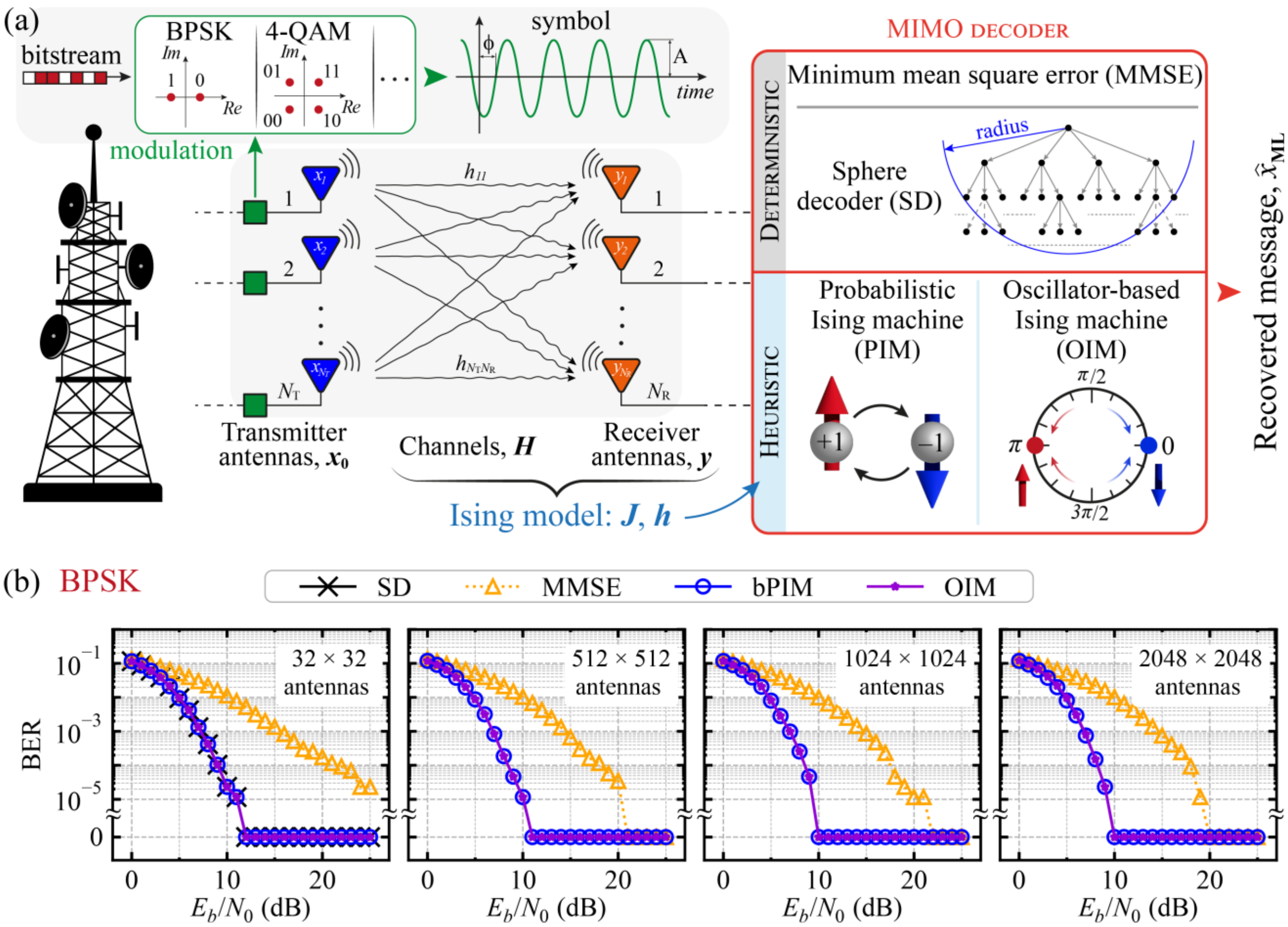

**Fig. 1**. (a) Illustrative scheme of the MIMO pipeline. At the top of the panel, the incoming bitstream is modulated into symbols according to minimum bit-error mapping over the chosen constellation (BPSK, 4-QAM, etc.). The message symbols $\boldsymbol{x}_0$, distributed across $N_T$ antennas, are transmitted through a channel $\boldsymbol{H}$ to $N_R$ antennas, resulting in the received symbols $\boldsymbol{y}$, affected by noise. To recover the original message, MIMO solvers, either with deterministic techniques (MMSE or SD) or heuristic ones (IMs), are used. For IMs, the Ising model ($\boldsymbol{J}$ and $\boldsymbol{h}$) is obtained directly from the channel matrix and the received message. (b) Deterministic and heuristic algorithm performance in terms of BER as a function of the bit energy-to-noise density ratio on BPSK instances of different sizes. The transmitted message is 86016 bits long. The black crosses show the theoretically optimal BER obtained through SD with infinite radius, only present in the 32×32 instances due to the long computation times. Both bPIM (in blue) and OIM (in purple) outperform MMSE (in yellow) and match SD for all $E_b/N_0$ values and all sizes. The zero error is due to the finite number of bits transmitted and provides an upper boundary for the true BER of less than $1.16 \cdot 10^{-5}$ with 95% confidence.

The MIMO problem consists in recovering the ML estimation of the original message $\hat{\boldsymbol{x}}_{ML} = \arg\min_{x \in \mathcal{S}^N} \| \boldsymbol{y} - \boldsymbol{H}\boldsymbol{x} \|^2$. Unless the transmission noise is high enough to push the received signal closer to another valid signal, the ground state $\hat{\boldsymbol{x}}_{ML}$ is usually equal to the original message $\boldsymbol{x}_0$ and finding it ensures the minimization of the BER. This can also be formulated as a Hamiltonian in the $\mathcal{S}^N$ space

$$\mathcal{H}(\boldsymbol{x}) = \|\boldsymbol{y} - \boldsymbol{H}\boldsymbol{x}\|^2. \tag{1}$$

In this formulation, the ground state of the Hamiltonian $\boldsymbol{x}_{GS}$ coincides with $\hat{\boldsymbol{x}}_{ML}$. The industry standard for MIMO solvers is the MMSE algorithm that estimates $\hat{\boldsymbol{x}}_{MMSE} = (\boldsymbol{H}^\dagger \boldsymbol{H} + \boldsymbol{I}\,\sigma^2/E_s)^{-1}\boldsymbol{H}^\dagger \boldsymbol{y}$ [8], where $\boldsymbol{I}$ is the identity matrix, and $\boldsymbol{H}^\dagger$ is the complex conjugate of the channel matrix. Its short computation time with polynomial complexity $O(N^3)$ makes it ideal for the low latency requirements of long-term evolution (LTE) and 5G communication at the cost of less accuracy than SD. On the other hand, the estimation with SD is deterministic and computes the optimal $\hat{\boldsymbol{x}}_{ML}$ at an exponential computational cost $O(M^N)$ as the size of the MIMO problem grows [6], [7].

To employ IMs to solve the MIMO problem heuristically, it is necessary to map the energy functional $\mathcal{H}(\boldsymbol{x})\colon \mathcal{S}^N \to \mathbb{R}$ to the Ising Hamiltonian $\mathcal{H}_b(\boldsymbol{s})\colon \{-1,+1\}^N \to \mathbb{R}$, with $\boldsymbol{s}$ being the binary spin state, and $\mathcal{H}_b$ being defined as

$$\mathcal{H}_b(\boldsymbol{s}) = -\frac{1}{2}\sum_{i,j} J_{b,ij}\, s_i s_j - \sum_i h_{b,i}\, s_i, \tag{2}$$

where $J_{b,ij}$ and $h_{b,i}$ are elements of the interaction matrix $\boldsymbol{J}_b$, and the bias matrix $\boldsymbol{h}_b$, respectively. These matrices are obtained by expanding the square of the real-valued expression of (1) and isolating the quadratic terms and the linear terms. The interaction matrix $\boldsymbol{J}_{\mathrm{b}}$ is entirely dependent on the channel matrix $\boldsymbol{H}$, while the bias matrix $\boldsymbol{h}_{\mathrm{b}}$ depends on both the channel matrix $\boldsymbol{H}$ and the received signal $\boldsymbol{y}$ (more details are provided in the Appendix A). Given $B = \log_2 \sqrt{M}$, the conversion from $\boldsymbol{s}$ to the real-valued symbol vector $\widetilde{\boldsymbol{x}}$ is achieved by considering the $B$-element vector $\boldsymbol{v} = [2^{-i}]_{i=1}^{B}$ and the $2N \times (2N \times B)$ transformation matrix $\boldsymbol{T} = \sqrt{M}\, \boldsymbol{v} \otimes \boldsymbol{I}$, where $\otimes$ is the Kronecker product, which results in $\widetilde{\boldsymbol{x}} = \boldsymbol{Ts}$ [14]. Starting from a symbol vector $\boldsymbol{x}$, its corresponding binary spin state $\boldsymbol{s}$ is obtained by algorithmically binarizing the constellation points of each symbol according to their coordinates. For BPSK, this means that $\boldsymbol{x} = \boldsymbol{s}$, while for QAM constellations, for an $N \times N$ instance, the size of $\boldsymbol{s}$ is $2N \log_2 \sqrt{M} = N \log_2 M$ .

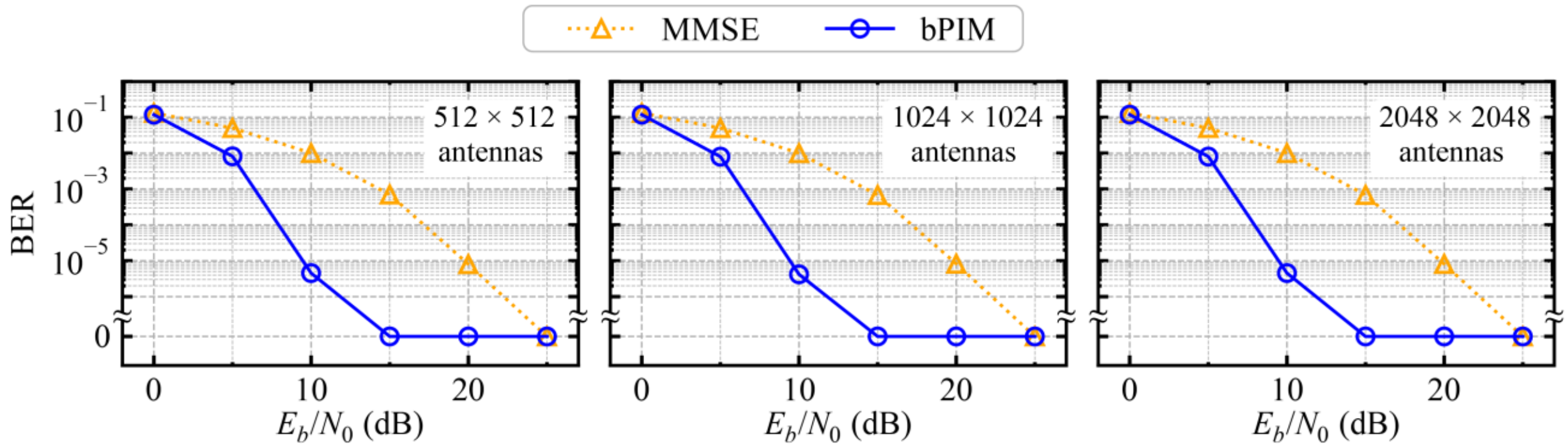

**Fig. 2**. Comparison of performance in terms of BER as a function of the bit energy-to-noise density ratio on BPSK instances of with 512, 1024 and 2048 antennas. The transmitted message consists of 8601600 bits. The bPIM (in blue) consistently outperform MMSE (in yellow) for all $E_b/N_0$ values and all the antenna sizes. The upper bound for the real BER of the BER = 0 noise values is $3.48 \cdot 10^{-7}$ with 95% confidence.

By finding the ground state $\boldsymbol{s}_{\mathrm{GS}}$ of (2), it is then possible to recover $\boldsymbol{x}_{\mathrm{GS}}$ and, thus, $\widehat{\boldsymbol{x}}_{\mathrm{ML}}$. The complete derivation of both $\boldsymbol{x} \rightarrow \boldsymbol{s}$ and $\boldsymbol{s} \rightarrow \boldsymbol{x}$ conversions is also reported in Appendix A.

## III. Probabilistic and Oscillator-Based Ising Machines

As a first step, the two heuristic approaches used to estimate $\widehat{\boldsymbol{x}}_{\mathrm{ML}}$ are the PIM and the OIM. The two methods were chosen as representatives of two very distinct classes of heuristic IM solvers, as the PIM is a discrete-time IM with discretized variables, while the OIM is an integrated-time IM with continuous variables. Their potential hardware implementations rely on complementary working principles [24], which makes them valid representatives to show the potential of IMs for MIMO decoding.

A PIM is a Markov-chain Monte Carlo (MCMC) algorithm that uses Gibbs sampling as its update kernel [25]–[28]. Each spin of the Ising model is a probabilistic bit (p-bit) that is updated in every algorithm iteration with update equation $s_i \leftarrow \mathrm{sgn}(\mathrm{rand}(-1, +1) + \tanh(\beta I_i(\boldsymbol{s})))$. Here, $\beta$ represents the control parameter of the PIM and acts as an inverse temperature, while $I_i(\boldsymbol{s})$ is the local field that acts on a specific p-bit and is given by $I_i(\boldsymbol{s}) = \sum_j J_{\mathrm{b},ij} s_j + h_{\mathrm{b},i}$ [25].

The OIM leverages the phase dynamics $\phi_i$ of coupled oscillators to minimize (2) using the Kuramoto model [29]–[32] (more details are presented in Appendix B). The control parameters of the OIM are $K$, $S$, and $T$ (see (11) in Appendix B). The $K$ parameter governs the coupling, $S$ the phase binarization, and $T$ the noise intensity.

To evaluate the performance of the IMs we have used the BER as a function of the bit energy-to-noise density ratio $E_b/N_0$, which is a well-established metric for the ML MIMO problem. In order to maintain the same BER resolution across all the problem sizes $N \times N$ shown, the number of bits transmitted was kept fixed to 86016. A total of $6144/N$ channels per size $N$, each tested with 14 different randomly generated messages, were considered.

Fig. 1(b) shows the performance of the bPIM and OIM algorithms compared with MMSE for BPSK modulation and sizes 32×32, 512×512, 1024×1024, and 2048×2048. The 32×32 configuration is also compared with results of the SD. The MMSE, in yellow, shows consistent performance across all sizes. The SD, black crosses in the figure, represents the theoretically optimal BER and it is only present for the 32×32 panel, due to the exponentially longer run times for larger sizes ($O(M^N)$). Both bPIM (in blue) and OIM (in magenta) are able to match the SD BER for the smallest size, and perform identically for the larger ones, suggesting they are still able to potentially reach the Ising ground states. Both paradigms operate using an annealing schedule, with $\beta$ linearly increasing for the bPIM and $T$ linearly decreasing for the OIM (while keeping $K$ and $S$ fixed). The parameter setting problem for these realizations of PIMs and OIM consists in identifying a common prescription to set the solver parameters ($\beta_{\max}$, $T_{\max}, K$, $S$) for a given MIMO problem instance [33]. As detailed in Appendix D, we identified a fixed parameter setting strategy consisting in using an analytical scaling relation for the optimal $\beta_{\max}$, $K$, and $S$, estimated empirically as $cN^{-2/3}$ for BPSK, where the $c$ constant was chosen as $c = \sqrt{3}$ for $\beta_{\max}$, $c = 3.5$ for $K$, and $c = 1.3$ for S, while $T_{\max} = 30$. A single solving attempt uses independently parallelized 64 replicas of the system ($R = 64$) for both bPIM and OIM, each consisting of 100 iterations ($N_{\mathrm{it}} = 100$), where an iteration is defined as a full system sweep for the bPIM and an integration step for the OIM (with $\Delta t = 0.01$ in dimensionless time).

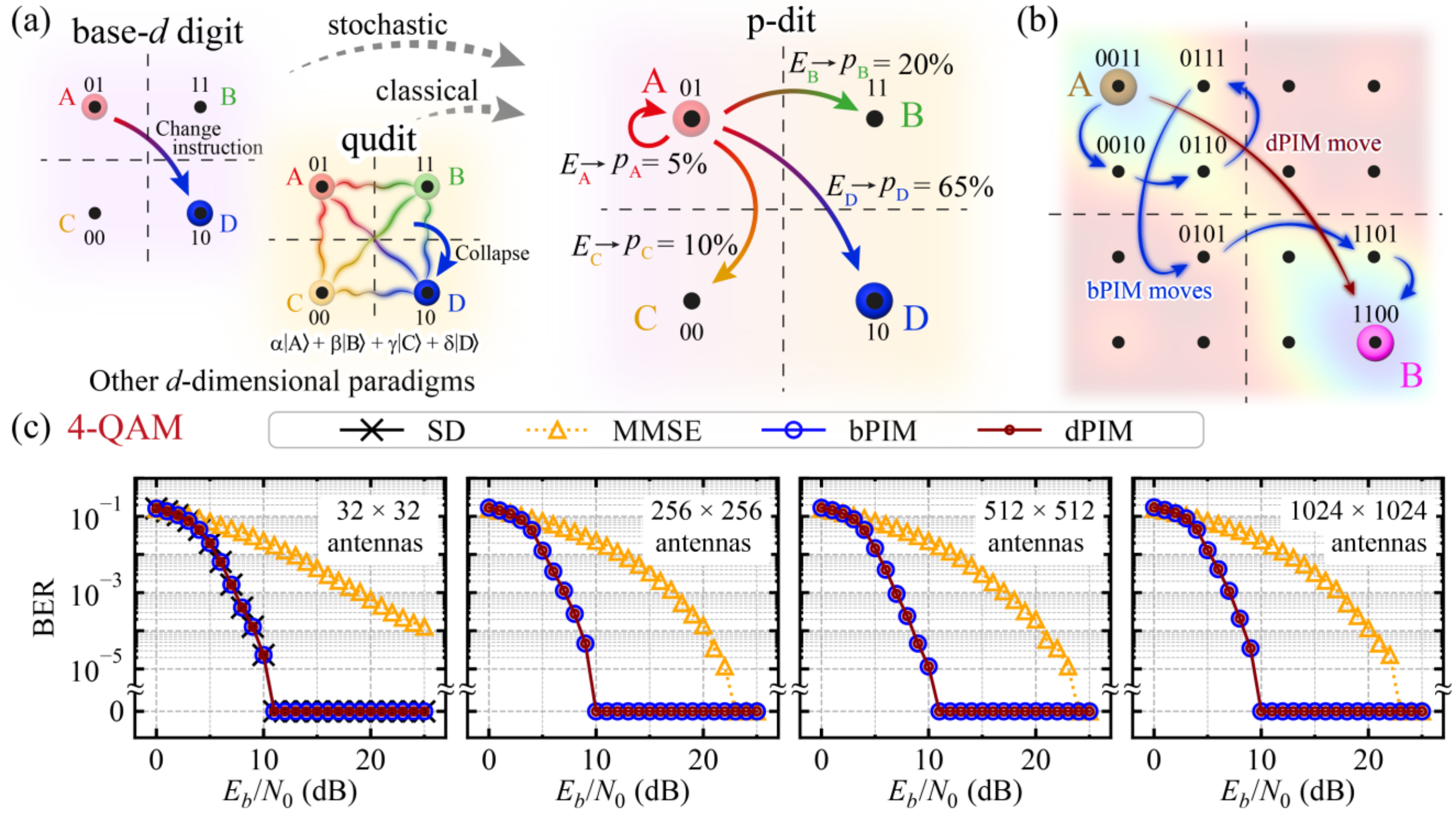

**Fig. 3**. (a) Illustrative representation of deterministic base-$d$ digits, quantum $d$-level systems (qudits), and $d$-dimensional probabilistic bits (p-dits) for $M$-QAM MIMO, the former two are included to better contextualize the latter. The base-$d$ bit evolves deterministically according to an algorithm; the qudit occupies a coherent superposition of some orthonormal basis states (represented here by the constellation symbols, with the state amplitudes $\alpha$, $\beta$, $\gamma$, $\delta$ being complex), and collapses probabilistically to one of them in the course of quantum measurement; the p-dit is a classical variable that evolves stochastically according to an energy functional. (b) Difference in energy landscape exploration between a p-bit and a p-dit. The former requires multiple low-probability moves to reach distant lower minima, while the latter requires a single evaluation and move. (c) Comparison of performance in terms of BER as a function of the bit energy-to-noise density ratio on 4-QAM instances of different sizes. The transmitted message is 86016 bits long. The black crosses show the theoretically optimal BER obtained through SD with infinite radius, only present in the 32×32 instances due to the long computation times. The two PIM paradigms, p-bits (in blue) and p-dits (in red) have matching performance for all sizes, and to SD for the smallest size, vastly outperforming MMSE (in yellow).

For the chosen bitstream length, the observed BER of zero in Fig. 1(b) provides an estimate of the upper bound for the true BER of $1.16 \cdot 10^{-5}$ with 95% confidence ($BER_{\text{up}} = -\ln(0.05)/N_{\text{bit}}$). Appendix E expands these results with a massive comparison for $N \times N$ ranging from 8×8 to 2048×2048.

To evaluate the compatibility of the bPIM with 6G reliability we performed additional tests for 512×512, 1024×1024, and 2048×2048 instances on a total of 8601600 bits using the same annealing scheme, resulting again in an observed BER of zero at $E_b/N_0$ higher than 10, as shown in Fig. 2. For these largest instances, the upper bound for the real BER is $3.48 \cdot 10^{-7}$ with 95% confidence. The main conclusion here is that, for BPSK modulation, different paradigms of IMs can estimate the $\hat{\boldsymbol{x}}_{\text{ML}}$ with the same accuracy as SD and perform systematically better than MMSE for MIMO instances with number of antennas at least up to 2048×2048. The main impact of this result is in MIMO applications where low signal-to-noise ratio conditions are likely or possible (like in satellite communications) and a more robust modulation scheme is needed [34], [35]. In the following, for decoding $M$-QAM modulations, we will focus on PIMs as they are a more mature paradigm from a technological point of view.

## IV. Modulations with High Spectral Efficiency and Generalized P-dit Mapping

Fig. 1(b) shows that an IM paradigm like the bPIM exhibits remarkable performance on BPSK MIMO instances with a very compact encoding. As we move to larger constellations with $M$-QAM modulations, however, the traditional binary Ising encoding requires a larger number of spins ($N \log_2 M$), as a single bit of information is not enough to uniquely identify a specific constellation point. This makes MIMO particularly suitable for the recently introduced generalized PIM paradigm, using probabilistic $d$-dimensional variables (p-dits) to represent the symbols [19].

The idea of p-dit, and its comparison with its deterministic and quantum counterpart to better contextualize it, is illustrated in Fig. 3(a) in a 4-QAM MIMO example. In general, a p-dit is a multi-dimensional probabilistic variable with an arbitrary number of values in each dimension that can be understood as a quantum-inspired computational unit. A standard digital representation of a multi-level state (a base-$d$ digit), like a 4-QAM constellation, as in Fig. 3(a) left, behaves deterministically and its dynamics are dictated by an algorithm. A quantum $d$-level system (a qudit), shown adjacent, exists in a coherent superposition of orthonormal basis states (constellation symbols), and collapses to one of them during measurement, with probability proportional to the squared modulus of the corresponding basis-state amplitude (i.e., coefficient of the corresponding basis vector). A p-dit stands in the middle between a base-$d$ digit and a qudit and fluctuates between its allowed states with a probability dictated by its energy functional. The right panel of Fig. 3(a) shows an example of a four-dimensional p-dit, with allowed states of $X = \{A, B, C, D\}$. As the dPIM is an MCMC using Gibbs sampling, the probability of a p-dit switching to any state ultimately does not depend on its current state. Without loss of generality, the probability of switching a given four-dimensional p-dit of a system from any state to state $A$ is

$$p_A = \frac{e^{-\beta E_A}}{e^{-\beta E_A} + e^{-\beta E_B} + e^{-\beta E_C} + e^{-\beta E_D}} , \quad (3)$$

where $E_A$ is the energy of the entire system if the currently considered p-dit is in state $A$. Additional details on the probability kernel of the dPIM are provided in Appendix C. The MIMO Hamiltonian for the dPIM does not require a transformation matrix $\boldsymbol{T}$ and can be written as

$$\mathcal{H}_{\mathrm{d}}(\boldsymbol{d}) = -\left(\sum_{a=1}^{2}\sum_{i=1}^{N} h_{\mathrm{d},i}^{a} d_i^a + \frac{1}{2}\sum_{a=1}^{2}\sum_{b=1}^{2}\sum_{i=1}^{N}\sum_{j=1}^{N} J_{\mathrm{d},ij}^{ab} d_i^a d_j^b\right), \quad (4)$$

where $\boldsymbol{d}$ is the p-dit symbol vector, of shape $N \times 2$, while $\boldsymbol{J}_{\mathrm{d}}$ and $\boldsymbol{h}_{\mathrm{d}}$ have, respectively, shape $N \times N \times 2 \times 2$ and $N \times 2$. The details on the mapping from (1) to the dPIM $\boldsymbol{J}_{\mathrm{d}}$ and $\boldsymbol{h}_{\mathrm{d}}$ are provided in Appendix C. The difference in energy of the $i^{\mathrm{th}}$ symbol switching from its current value $x_0$, mapped as $d_0$, to a new value $x_1$ in $\mathcal{S}$, mapped as $d_1$, is given by

$$\Delta E_{i,d_0\to d_1} = \sum_{a=1}^{2}\left[(d_1^a - d_0^a) I_i^a - \frac{1}{2}\sum_{b=1}^{2} J_{\mathrm{d},ii}^{ab}(d_1^a d_1^b - 2 d_1^a d_0^b + d_0^a d_0^b)\right], \quad (5)$$

where $I_i^a = h_{\mathrm{d},i}^a + \sum_j^N \sum_b^2 J_{\mathrm{d},ij}^{ab}\, d_j^b$. The main difference between the bPIM and the dPIM, which makes the latter more favorable for a problem like MIMO is in how they explore the energy landscape for each symbol variable. As illustrated by Fig. 3(b), a p-bit, due to its binary encoding that requires multiple spins per symbol, requires several low-probability moves to traverse an energy landscape that presents two energy minima at a large Hamming distance. Conversely, a p-dit evaluates the full energy spectrum during its update for a given symbol, requiring a single high-probability move.

Fig. 3(c) shows a comparison of the performance of the bPIM and dPIM for 4-QAM instances for $N = 32\times32$, 256×256, 512×512, and 1024×1024 with the same BER resolution as Fig. 1(b). The SD baseline is present only in the first panel due to computational limitations. Appendix E expands these results with a systematic comparison of sizes ranging from 8×8 to 1024×1024. The dPIM (in red) and bPIM (in blue) have matching performance for all considered problem sizes and signal-to-noise ratio values. Both algorithms use $R = 64$ and $N_{\mathrm{it}} = 100$. As detailed in Appendix D, the annealing parameter follows a scaling relation for the bPIM given by $\beta = 13/(N\sqrt{M})$, while for the dPIM it was empirically estimated as $\beta = \sqrt{2}\, N^{-4/5}$. It is noteworthy how the heuristically determined optimal parameters for the bPIM appear to be dependent on the constellation size, while the dPIM ones do not. This can be attributed to the fact that the size of the $\boldsymbol{J}_{\mathrm{b}}$ matrix for the bPIM depends on $M$, while the dPIM interaction matrix $\boldsymbol{J}_{\mathrm{d}}$ only depends on $N$.

For the 32×32 instances, whose search space is still computationally tractable with SD, both heuristic algorithms exhibit evidence of optimal performance as they match the theoretical ML BER (black crosses). In 4-QAM, the MMSE (in yellow) is vastly outperformed by both dPIM and bPIM for all tested sizes within the BER resolution limits of the number of bits transmitted. It is worth noting that the two algorithms would eventually exhibit an error floor if the number of transmitted bits was increased, with dPIM's one expected to be lower. For higher QAM, the bPIM performance degrades rapidly, most likely due to the exponential increase of the spin configuration space $\left(2^{N\log_2 M}\right)$ and, consequently, of local minima. This increases exponentially the number of statistical fluctuations required to overcome the entropic barrier, justifying the degradation in performance.

In addition to the performance gain, there is also another advantage in using the p-dit formulation rather than the p-bit one for an $M$-QAM MIMO IM implementation: the interaction matrix $\boldsymbol{J}_{\mathrm{d}}$ does not depend on the $M$-QAM used, but only on the channel matrix $\boldsymbol{H}$. This eliminates the need to recalculate the $\boldsymbol{J}_{\mathrm{d}}$ if the $M$-QAM modulation changes dynamically, limiting it only to when the wireless interference characteristics have changed. This is relevant for applications in which physical conditions change slowly (quasi-static environments) but sudden drops in the channel quality indicator (CQI) or spikes in network load can require a change in modulation.

To analyze the scalability of the dPIM, a systematic test was carried out on instances with larger modulations and for several choices of $N$. Fig. 4 summarizes the ones achieved for $M$-QAM modulation $M = 16$, 64, 256 and number of antennas $N \times N$ ranging from 8×8 to 256×256. Each panel of Fig. 4 shows an $M, N$ combination, with different rows/columns referring to different sizes/modulations, respectively. In all panels, the total number of transmitted bits was kept consistent with Fig. 1(b) and Fig. 2(b), at 86016. The ZF and MMSE are shown in grey and yellow, respectively, as a baseline reference. The ZF exhibits worse performance compared to MMSE, at a gain of requiring no prior information on the signal-to-noise ratio affecting the transmission. In the panels framed in green, the SD is represented as black crosses and shows the theoretically optimal decoding for computationally feasible problem sizes.

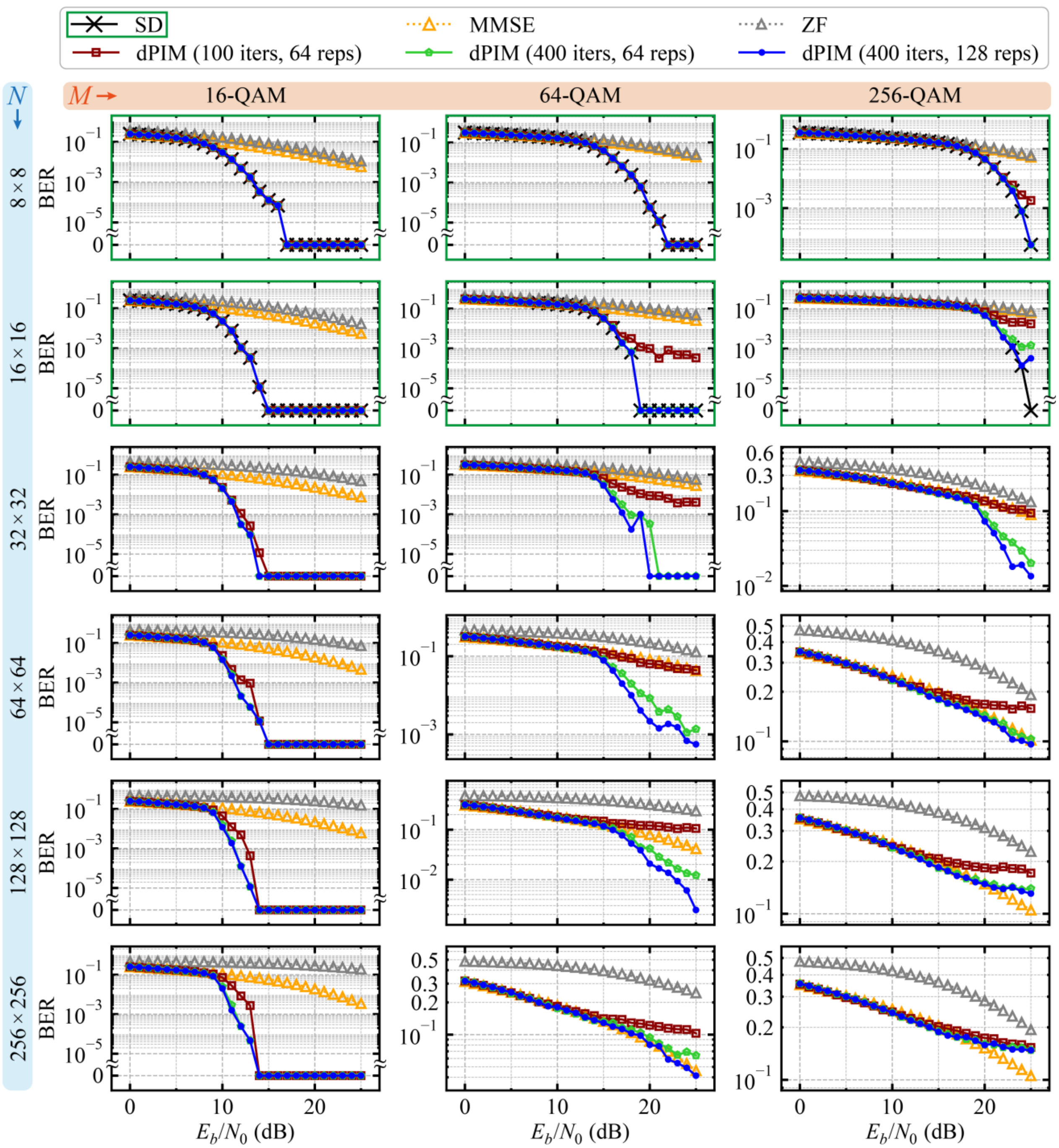

**Fig. 4**. Comparison of performance in terms of BER as a function of the bit energy-to-noise density ratio on instances of different sizes (different rows) and modulation (different columns). The transmitted message is 86016 bits long. The black crosses show the theoretically optimal BER obtained through SD with infinite radius, only present on panels framed in green. The ZF (in grey) and MMSE (in yellow) are shown as a baseline comparison. The dPIM performance is presented for different combinations $(N_{\text{it}}, R)$ of number of iterations $N_{\text{it}}$ and replicas $R$. Respectively, (100,64),(400,64), and (400,128) are shown in red, green, and blue. Note the difference in BER scales in each plot.

The dPIM performance is shown for different combinations of the number of replicas and iterations. The red curves show the performance with the same parameters as those used for Fig. 2(b), $R = 64, N_{\text{it}} = 100$. The effect of a longer schedule is illustrated by the curves in green, with $R = 64, N_{\text{it}} = 400$. The gain from using more replicas is shown by the blue curves, with $R = 128, N_{\text{it}} = 400$.

The dPIM shows good performance across all sizes and modulations. While it was too computationally demanding to obtain the SD BER for all $M, N$ combinations, the dPIM closely matches the SD performance for the smaller problem sizes. The dPIM outperforms ZF for all $E_b/N_0$ values tested, and for all modulations and problem sizes considered, with the less computationally intensive schedule parameters ($R = 64, N_{\text{it}} = 100$). Notably, it also does not require signal-to-noise ratio information.

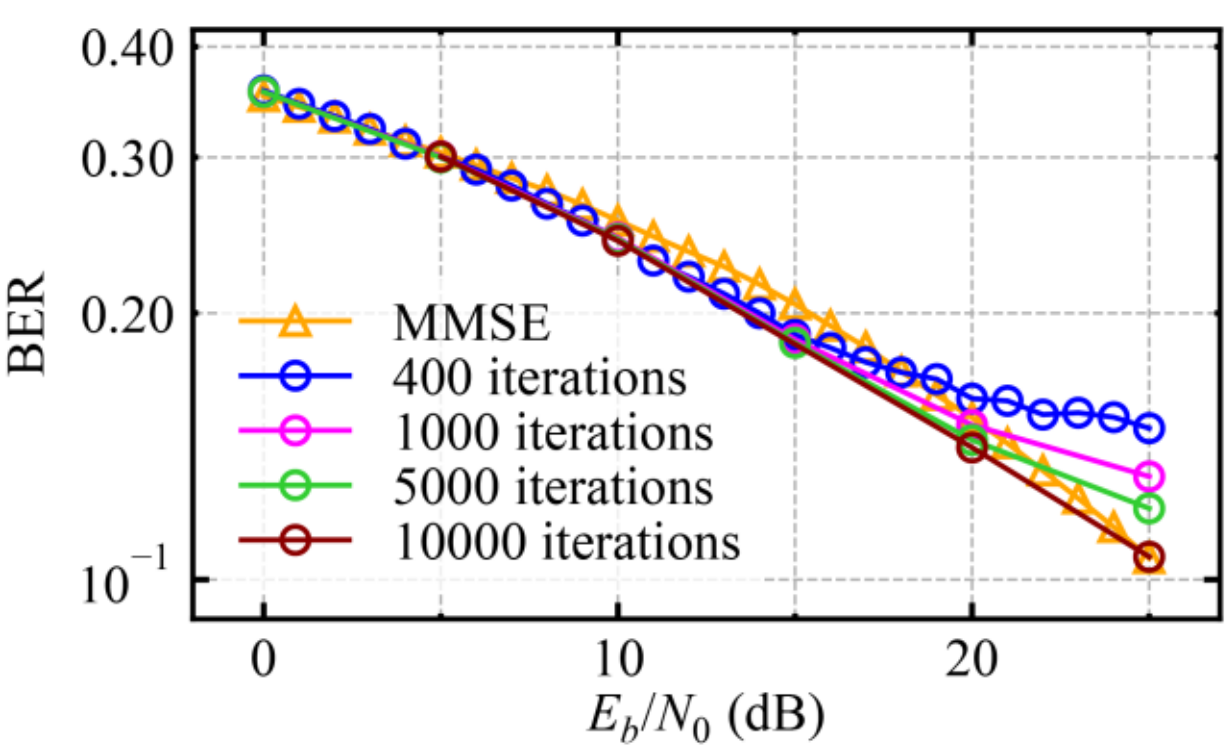

**Fig. 5**. Comparison of the dPIM performance in terms of BER for 256-QAM and $256 \times 256$ antennas as a function of the bit energy-to-noise density ratio with 64 replicas and 400 (in blue), 1000 (in magenta), 5000 (in green), and 10000 (in red) iterations. The transmitted message is 86016 bits long. MMSE is in yellow.

Moreover, for 16-QAM instances, the dPIM outperforms MMSE for all $E_b/N_0$values. The dPIM error floor can be estimated to be lower than $1.16 \cdot 10^{-5}$ with 95% confidence. For those instances, the length of the schedule and the number of replicas show only marginal impact on the performance with the BER resolution considered.

For 64-QAM, longer annealing is required to match SD for $N = 16$, and to outperform MMSE for $N = 64$. With 256 antennas, the dPIM is able to match MMSE with the $R = 128, N_{\text{it}} = 400$ schedule. For 256-QAM, the dPIM greatly outperforms MMSE for 16 and 32 antennas, and closely matches the SD BER. For 64 antennas, the $R = 128, N_{\text{it}} = 400$ dPIM schedule matches MMSE performance for the entire noise range tested, while for 256 antennas MMSE shows better BER for $E_b/N_0 > 18$. Moreover, for $R = 64$, when the number of iterations is increased further, the dPIM eventually matches MMSE, as shown in Fig. 5.

## IV. Conclusion

The presented results suggest that the potential impact of IM-based heuristic MIMO solvers extends well beyond proof-of-principle demonstrations. For BPSK, both bPIM and OIMs achieve SD-level detection when ML optimization is computationally tractable and achieve better BER performance compared to MMSE with only 100 iterations, even when the system is scaled to XL-MIMO sizes, up to 2048×2048 antennas. However, for $M$-QAM at large $M$, the key bottleneck for p-bits is that the binary encoding of $M$-QAM expands the dimension of the spin configuration space, whereas the p-dit encoding preserves the native structure of QAM symbols and thereby reduces the error floor given by the entropic barrier. This encoding advantage is what enables the enhanced performance of the dPIM from 4-QAM to 256-QAM across a broad range of antenna dimensions.

Beyond BER performance, the p-dit formulation has an important architectural consequence. Because the dPIM coupling matrix is determined only by the channel matrix $\boldsymbol{H}$ rather than also depending on the modulation order, changes in constellation do not require rebuilding the full interaction matrix, but only the local field bias, which needs to be computed for every new transmission in any case. This makes adaptive modulation natively compatible with the proposed architecture and suggests a natural route toward reconfigurable IM-based receivers, in which spectral efficiency can be tuned directly according to channel conditions [36]. In fact, in modern wireless systems, adaptive modulation and coding are often employed, where the modulation order is selected based on channel quality feedback, typically evaluated at the subframe level (1 ms), as specified in the Technical Specification 3GPP TS 36.213 [37]. In this sense, p-dits can serve as a building block to improve heuristic detection accuracy while also providing an algorithmic interface between near-ML inference and link adaptation in future wireless systems.

The computational complexity of the bPIM and the dPIM paradigms is $O(N_{\text{iter}} N^2)$ as a function of problem size and number of iterations, with the bPIM having a larger prefactor due to the binary encoding of the symbol state. This computational cost is comparable with low complexity MMSE approaches [38]. While out of scope of the current proof-of-concept work, operational timing estimates on efficient implementations of the discussed solvers can be compatible with the required latency for real applications. For instance, a bPIM implementation on FPGA [16] totaled 0.47 ms for 100 sweeps on sparsified 64×64 BPSK MIMO instances. For higher modulation, the overhead from bPIM to dPIM on the sweep time would introduce at worst a factor of $\sim M$. It is reasonable to anticipate ASIC speedups and further algorithmic implementation bringing the timescales under a millisecond for XL-MIMO applications.

In conclusion, our results establish p-dit probabilistic Ising machines as a scalable and reconfigurable platform for XL-MIMO detection, and as a particularly promising foundation for future adaptive-modulation receivers, and will serve as foundation for the design of physical embodiments of the dPIM where stochasticity is natively available [39], [40], including future qudit processors [41].

## Appendix A
### Traditional Ising mapping of the MIMO problem

To map the MIMO energy functional of (1), the complex-valued equation has to be rewritten as a real-valued one. This can be achieved with the following transformation of the channel matrix $\boldsymbol{H}$, the received message $\boldsymbol{y}$ and symbol vector $\boldsymbol{x}$ [14]:

$$\widetilde{\boldsymbol{H}} = \begin{bmatrix} \Re(\boldsymbol{H}) & -\Im(\boldsymbol{H}) \\ \Im(\boldsymbol{H}) & \Re(\boldsymbol{H}) \end{bmatrix}, \quad \widetilde{\boldsymbol{y}} = \begin{bmatrix} \Re(\boldsymbol{y}) \\ \Im(\boldsymbol{y}) \end{bmatrix}, \quad \widetilde{\boldsymbol{x}} = \begin{bmatrix} \Re(\boldsymbol{x}) \\ \Im(\boldsymbol{x}) \end{bmatrix}, \tag{6}$$

with $\Re$ and $\Im$ being the real and imaginary part operators. The transformation guarantees that the ground state of $\widetilde{\mathcal{H}}(\widetilde{\boldsymbol{x}}) = \|\widetilde{\boldsymbol{y}} - \widetilde{\boldsymbol{H}}\widetilde{\boldsymbol{x}}\|^2$ is the real-valued equivalent of $\boldsymbol{x}_{\mathrm{GS}}$. For the BPSK case, in which the constellation symbols are real-valued, the real-valued channel can be defined as

$$\widetilde{\boldsymbol{H}}_{\mathrm{BPSK}} = \begin{bmatrix} \Re(\boldsymbol{H}) \\ \Im(\boldsymbol{H}) \end{bmatrix}. \tag{7}$$

The real-valued Hamiltonian can be expanded and separated into three components as follows:

$$\begin{aligned} \widetilde{\mathcal{H}}(\widetilde{\boldsymbol{x}}) &= \|\widetilde{\boldsymbol{y}} - \widetilde{\boldsymbol{H}}\widetilde{\boldsymbol{x}}\|^2, \\ \widetilde{\mathcal{H}}(\widetilde{\boldsymbol{x}}) &= (\widetilde{\boldsymbol{y}} - \widetilde{\boldsymbol{H}}\widetilde{\boldsymbol{x}})^\dagger (\widetilde{\boldsymbol{y}} - \widetilde{\boldsymbol{H}}\widetilde{\boldsymbol{x}}), \\ \widetilde{\mathcal{H}}(\widetilde{\boldsymbol{x}}) &= \underbrace{\widetilde{\boldsymbol{x}}\widetilde{\boldsymbol{H}}^\dagger\widetilde{\boldsymbol{H}}\widetilde{\boldsymbol{x}}}_{\text{Quadratic}} - \underbrace{2\widetilde{\boldsymbol{y}}^\dagger\widetilde{\boldsymbol{H}}\widetilde{\boldsymbol{x}}}_{\text{Linear}} + \underbrace{\widetilde{\boldsymbol{y}}^\dagger\widetilde{\boldsymbol{y}}}_{\text{Constant}}, \end{aligned} \tag{8}$$

Defining the constant $B = \log_2 \sqrt{M}$, and the $B$-dimensional vector $\boldsymbol{v} = [2^{-i}]_{i=1}^{B}$, we obtain the transformation matrix defined in the main text as $\boldsymbol{T} = \sqrt{M}\, \boldsymbol{v} \otimes \boldsymbol{I}$ to convert $\widetilde{\boldsymbol{x}} = \boldsymbol{T}\boldsymbol{s}$. Neglecting the constant terms (irrelevant for optimization), one can write the Ising Hamiltonian as

$$\begin{aligned} \mathcal{H}_I(\boldsymbol{s}) = 2\boldsymbol{s}^\dagger \left(\boldsymbol{T}^\dagger\widetilde{\boldsymbol{H}}^\dagger\widetilde{\boldsymbol{H}}\boldsymbol{T} \circ (\boldsymbol{1} - \boldsymbol{I})\right)\boldsymbol{s} \\ - 2\widetilde{\boldsymbol{y}}^\dagger\widetilde{\boldsymbol{H}}\boldsymbol{T}\boldsymbol{s}, \end{aligned} \tag{9}$$

$$\mathcal{H}_I(\boldsymbol{s}) = -\boldsymbol{s}^\dagger \frac{\boldsymbol{J}}{2}\boldsymbol{s} - \boldsymbol{h}^\dagger\boldsymbol{s}, \tag{10}$$

where $\boldsymbol{1}$ is a square matrix of ones, and $\circ$ is the Hadamard (elementwise) product.

## APPENDIX B
## OSCILLATOR-BASED ISING MACHINE IMPLEMENTATION

The OIM, an IM paradigm in which the spin variables are encoded as the phases of oscillators, is implemented using the Kuramoto model [29]–[31], with the additional assumption that all oscillators have the same natural frequency to ease synchronization. For each spin variable $s_i$, the corresponding phase $\phi_i(t)$ evolves continuously in time and its dynamics governed by

$$\dot{\phi}_i(t) = -K\, \mathcal{C}_i(t) - S\, \mathcal{B}_i(t) + T\, \xi(t), \tag{11}$$

where $\mathcal{C}_i(t)$ is the coupling function of $\phi_i$, $\mathcal{B}_i(t)$ is the binarization function of $\phi_i$, and $\xi(t)$ the additive white Gaussian noise function. The $K$ parameter governs the coupling, the $S$ parameter the phase binarization, the $T$ parameter the noise intensity. The coupling and binarization function are given by

$$\begin{aligned} \mathcal{C}_i(t) = \sum_j J_{\mathrm{b},ij} \tanh\left(10 \sin\left(\phi_i(t) - \phi_j(t)\right)\right) \\ + h_{\mathrm{b},i} \sin(-\phi_i(t)), \end{aligned} \tag{12}$$

$$\mathcal{B}_i(t) = \sin(2\phi_i(t)). \tag{13}$$

The dynamics for each oscillator are integrated with Heun's method using time step $\Delta t = 0.01$.

## APPENDIX C
## EXTENDED VARIABLES MAPPING OF THE MIMO PROBLEM

A 2-dimensional, $M$ state, p-dit variant can be used to represent each symbol $\boldsymbol{x}$. The first dimension of the $i^{\text{th}}$ p-dit $d_i^1$ holds one of $\sqrt{M}$ integers $\{\dots, -5, -3, -1, +1, +3, +5, \dots\}$ that correspond to the real portion of the signal. As $d_i^2$ holds the same possible values and corresponds to the imaginary portion of the signal, a total of $M$ states can be represented by a single p-dit, equivalent to the original symbol. A system of $N_{\mathrm{T}}$ p-dits fully describes the equation outlined in (1).

If we continue to use superscript values of 1 and 2 to indicate, respectively, the real and imaginary components of the channel $\boldsymbol{H}$ and received signal $\boldsymbol{y}$, we can define each dimension of the bias vector $\boldsymbol{h}_{\mathrm{d}}$, as follows

$$h_{\mathrm{d},i}^1 = 2\sum_k^N H_{ki}^1 y_k^1 + H_{ki}^2 y_k^2, \tag{14}$$

$$h_{\mathrm{d},i}^2 = 2\sum_k^N H_{ki}^1 y_k^2 - H_{ki}^2 y_k^1. \tag{15}$$

Each dimension of the coupling matrix $\boldsymbol{J}_{\mathrm{d}}$ describes the correlation between two pairs of p-dit dimensions with each element taking the value of

$$J_{\mathrm{d},ij}^{11} = J_{\mathrm{d},ij}^{22} = -2\sum_k^N \left(H_{ki}^1 H_{kj}^1 + H_{ki}^2 H_{kj}^2\right), \tag{16}$$

$$J_{\mathrm{d},ij}^{12} = -J_{\mathrm{d},ij}^{21} = -2\sum_k^N \left(-H_{ki}^1 H_{kj}^2 + H_{ki}^2 H_{kj}^1\right). \tag{17}$$

Note that the value of $\boldsymbol{J}_{\mathrm{d}}$ only depends on $\boldsymbol{H}$, while $\boldsymbol{h}_{\mathrm{d}}$ depends on $\boldsymbol{H}$ and $\boldsymbol{y}$. Moreover, we only have to explicitly store $\boldsymbol{J}_{\mathrm{d}}^{11}$ and $\boldsymbol{J}_{\mathrm{d}}^{12}$. The Ising Hamiltonian can then be defined as

$$\begin{aligned} \mathcal{H}_{\mathrm{d}}(\boldsymbol{d}) = -\Bigg(\sum_{a=1}^{2}\sum_{i=1}^{N} h_{\mathrm{d},i}^a d_i^a \\ + \frac{1}{2}\sum_{a=1}^{2}\sum_{b=1}^{2}\sum_{i=1}^{N}\sum_{j=1}^{N} J_{\mathrm{d},ij}^{ab} d_i^a d_j^b\Bigg). \end{aligned} \tag{18}$$

For simplicity we can define

$$I_i^a = h_{\mathrm{d},i}^a + \sum_j^N \sum_b^2 J_{\mathrm{d},ij}^{ab}\, d_j^b. \tag{19}$$

Then, the energy difference between any two values of a p-dit becomes

$$\begin{aligned} \Delta E_{i,d_0 \to d_1} = \sum_{a=1}^{2} \Bigg[ (d_1^a - d_0^a) I_i^a \\ - \frac{1}{2}\sum_{b=1}^{2} J_{\mathrm{d},ii}^{ab}\big(d_1^a d_1^b - 2 d_1^a d_0^b \\ + d_0^a d_0^b\big)\Bigg]. \end{aligned} \tag{20}$$

When a p-dit is updated, it can move to any of its $M$ possible states. If these are labeled as $\{A, B, C, \dots\}$ then, without loss of generality, each has a probability of

$$p_A = \frac{1}{1 + \exp(\beta\Delta E_{i,B\to A}) + \exp(\beta\Delta E_{i,C\to A}) + \cdots}. \tag{21}$$

## APPENDIX D
## PARAMETER OPTIMIZATION FOR BPIM, DPIM, AND OIM

We performed simulations for the various Ising machine (IM) paradigms using simulated annealing (SA) for a fixed number of iterations $N_{\mathrm{it}}$, here $N_{\mathrm{it}} = 100$, as a function of the paradigm temperature parameter. Using the normalized average final energy reached by the solver $\langle E_{\text{F}} \rangle$ as the metric of choice, we observe that different instances at the same modulation and size exhibit similar trends. Using the bPIM parameters as an example, for $\beta_{\max} \to 0$ the energy reaches the average energy of a random spin configuration; as $\beta_{\max}$ increases, the $\langle E_{\text{F}} \rangle$ sharply decreases until it reaches a minimum $\beta_O$; with $\beta_{\max} > \beta_O$, the $\langle E_{\text{F}} \rangle$ slowly rises as the system is frozen increasingly too quickly to escape shallow local minima. While this behavior appears to be universal, the value of the minimum $\beta_O$, which represents the most desirable choice for $\beta_{\max}$ to minimize the BER, changes drastically as a function of the number of antennas $N$ (for all considered paradigms) and the modulation $M$ (for the bPIM and the OIM). Note some acronyms and parameters are defined in the main text.

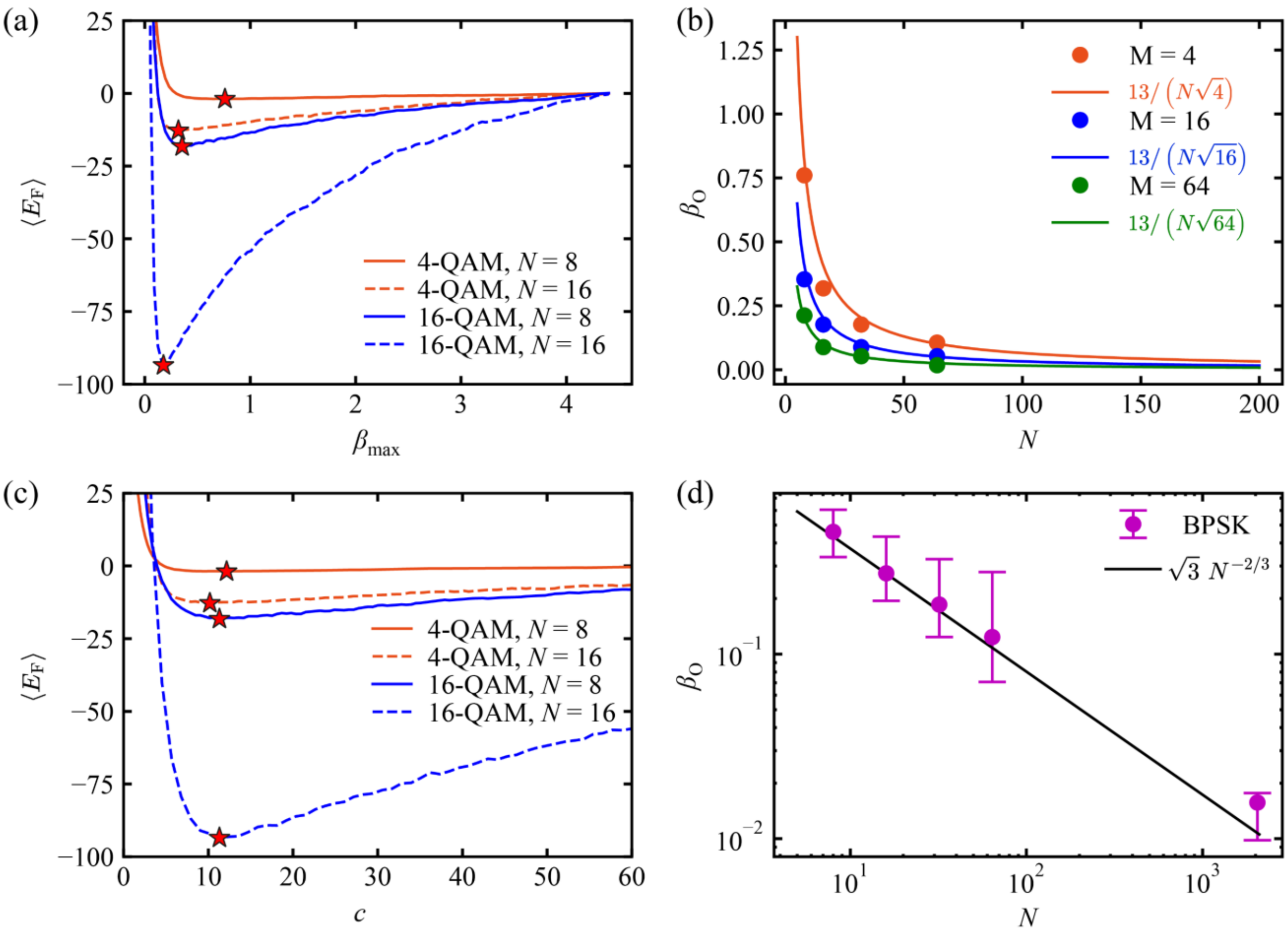

**Fig. 6**. (a) Average final energy as a function of the chosen $\beta_{\max}$ for SA schedules with $N_{\text{it}} = 100$. Each point is average across 3 values of $E_b/N_0$, 20 instances per $E_b/N_0$, and 100 trials per instance. The orange, blue lines show 4-QAM, 16-QAM performance, respectively, while full, dashed lines show 8×8, 16×16 size, respectively. (b) Optimal $\beta_{\max}$ as a function of $N$ for 4-QAM (in orange), 16-QAM (in blue), and 64-QAM (in green). Each point is the mean value between the 60 instances considered. The full lines represent the fitting with equation $\beta_{\max} = 13/N\sqrt{M}$. (c) Average final energy from (a) as a function of the normalizing parameter $c$, with $\beta_{\max}$ scaled by a factor $N\sqrt{M}$. The minima are aligned at $c = 13$. (d) Optimal $\beta_{\max}$ ($\beta_O$) as a function of $N$ in log-log scale for BPSK instances. Each point is the median between the 60 instances considered, the error bar showing the 25% and 75% quartile. The fitting line follows the equation $\beta_{\max} = \sqrt{3}N^{-2/3}$.

As an example, Fig. 6(a) shows the average final energy $\langle E_F \rangle$ of the bPIM as a function of the $\beta_{\max}$ for 4-QAM and 16-QAM, and sizes 8×8 and 16×16. For the same size, larger modulation results in the minimum being at a lower $\beta_{\max}$; similarly, for the same modulation, larger sizes result in lower $\beta_O$. The curves shown here are averaged across instances taken from signal-to-noise ratio values $E_b/N_0 = 3, 6, 9$ dB, for a total of 20 instances per $E_b/N_0$ value, and 100 solving tests per instance. By collecting the optimal $\beta_{\max}$ for several sizes and modulations, we can observe a clear trend as a function of $N$ and $M$, as shown by Fig. 6(b), that displays $\beta_O$ as function of $N$ for $M = 4, 16, 64$. Each point in the graph is obtained with the mean across the 60 instances considered. The optimal $\beta_O$ found is inversely proportional to $N$ and $\sqrt{M}$, as highlighted by the full lines representing the equation $c/(N\sqrt{M})$ with $c = 13$. Fig. 6(c) displays the same curve of Fig. 6(a) scaled by a factor $\propto N\sqrt{M}$. It is possible to observe the minimum $c$ for each curve being aligned at approximately $c = 13$. The parameter thus determined was also used to extrapolate the optimal $\beta_{\max}$ for 256-QAM.

While all $M$-QAM constellations have similar structure and encoding characteristics for bPIM and OIM, with the number of spins (and related sizes of the $\boldsymbol{J}_{\text{b}}$ and $\boldsymbol{h}_{\text{b}}$ matrices) being $2N \log_2 \sqrt{M}$, the BPSK is encoded in both bPIM and OIM with $N$ spins. This is due to the BPSK constellation symbols being real values, which results in $\tilde{\boldsymbol{x}} = \boldsymbol{x}$, as mentioned in Appendix A. Performing the same minimum analysis of Fig.

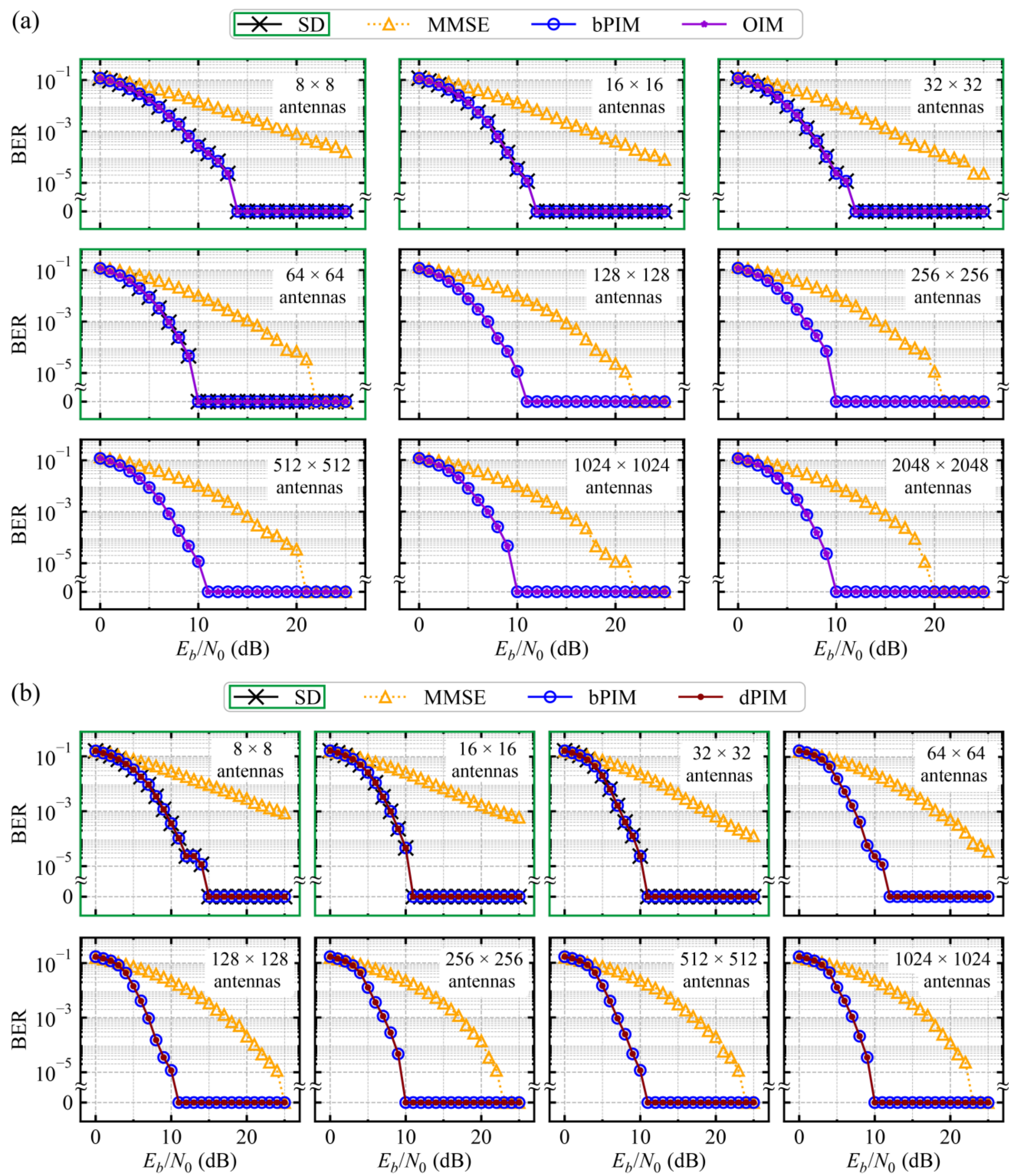

**Fig. 7**. Comparison of performance in terms of BER as a function of the bit energy-to-noise density ratio on instances of different sizes, ranging from 8 to 2048 antennas in powers-of-two steps for BPSK in (a), and from 8 to 1024 for 4-QAM in (b). The transmitted message is 86016 bits long. (a) Both bPIM (in blue) and OIM (in purple) match the theoretically optimal BER obtained through SD with infinite radius (black crosses) in the panels highlighted with a green frame. In addition, bPIM and OIM outperform MMSE (in yellow) for all $E_b/N_0$ values and all sizes. The upper bound for the BER for the noise values with BER = 0 is $1.16 \cdot 10^{-5}$ with 95% confidence. (b) The two PIM paradigms, p-bits (in blue) and p-dits (in red) have matching performance and outperform MMSE (in yellow) for all sizes and match SD (black crosses) in the panels with a green frame.

6(b) for BPSK shows a different trend compared to the $M$-QAM constellations, as shown in Fig. 6(d), in log-log scale to better appreciate points at very different sizes. While the $\beta_O \propto N^{-1}$ scaling applies to all $M$-QAM, the BPSK power law scaling appears to have an exponent greater than $-1$. Each point in the graph is obtained with the median across the 60 instances considered, the error bars being the 25% and 75% quartile. The full line represents the fitting, which results in $\beta_{\max} = \sqrt{3}N^{-2/3}$. This choice of the constant and exponent is able to capture even points at much larger $N$, like the 2048×2048 instances of Fig. 6(b).

The dPIM, as it encodes problems of all QAM using the same procedure and equal number of computational units, due to the flexible p-dit formulation, showed a scaling of $\beta_O$ independent of $M$, with the optimal fitting being $\beta_{\max} = \sqrt{2}N^{-4/5}$. The OIM, in which the temperature is represented by the ratio between the noise parameter $T$ and the interaction and binarization parameters $K$ and $S$, uses the same scaling law as the bPIM $\beta_{\max}$ with a different constant $c$, while the ratio between $K$ and $S$ is kept constant according to previous studies reported in literature [31].

## Appendix E
## BER performance on BPSK and 4-QAM instances scaling with number of antennas

Fig. 7 shows extended results on BPSK, in (a), and 4-QAM instances, in (b), with scaling number of antennas.